# Neutrino spin and dispersion in magnetized medium

P.A. Eminov [a,b]

[a] Moscow State University of Information Technologies, Radioengineering and Electronics

Russia, Moscow 107996, Ulitsa Strominka 20

[b] National Research University Higher School of Economics

Russia, Moscow 101000, Myasnitskaya Ulitsa 20

E-mail: peminov@mail.ru

**Abstract**

The full energy shift of a massive Dirac neutrino in magnetized electron-positron plasma was investigated using the Matsubara imaginary time and real time formalisms. The neutrino dispersion in the magnetized medium was analyzed as a function of the neutrino spin and mass. It was shown that in a super-strong magnetic field the CP-symmetric plasma contribution to the neutrino energy greatly exceeds the analogous correction in the field-free case. The contribution of plasma to the anomalous magnetic moment of a neutrino was obtained.

## 1. Introduction

In recent years, investigating neutrino properties in a dense medium at finite temperatures in the presence of external electromagnetic fields has become of considerable interest in the field of neutrino physics and astrophysics. A comprehensive review of previous studies related to collective neutrino-plasma interactions, as well as their possible astrophysical applications, is presented in [1-21]. Experimental observations of neutrino oscillations (see, e.g., [22-23]) have provided reliable evidence that neutrinos possess nonzero masses and that they mix. The energy dispersion relation for massless Dirac neutrinos in the presence of a constant magnetic field in plasma has been studied extensively in the literature [24-33].

Nevertheless, a calculation of the neutrino self-energy in a magnetized media that also accounts for the effects of the spin and mass of the neutrino has not been presented.

In this study, we shall concentrate on studying the full energy shift and anomalous magnetic moment (AMM) for a massive Dirac neutrino in a magnetized electron-positron plasma. The paper is organized as follows. In Sec.2, we discuss about the general properties of the neutrino self-energy in a magnetized medium and calculate the contributions from the weak charge current to the one-loop energy shift and anomalous magnetic moment of an electron neutrino using the Matsubara temperature Green's functions. In Section 3, we determine the contribution of the neutral weak current to the neutrino dispersion and AMM in a magnetized medium, and approximate formulas are derived for each flavor of the propagating neutrino.

In Sec. 4, we calculate the CP-symmetric plasma contribution of the neutrino dispersion relation in a strong magnetic field using a real time representation of the Green's function of an ideal electron-positron gas in a constant magnetic field. We also estimate numerical values of the expected contributions to the energy shift and AMM of a neutrino. Finally, we summarize our results in Sec. 5.

**2. Neutrino self-energy in a magnetized medium: imaginary time representation and charged weak current contribution**

The electron neutrino interacts with electrons both due to the neutral weak current and the charged weak current, whereas neutrinos of other types interact with electrons only through the neutral current. In the Feynman gauge, the three Feynman diagrams contribute to the neutrino self-energy in a magnetized medium: the bubble diagram with the W-boson, the bubble diagram with the charged scalar and the tadpole diagram with the Z-boson [27-29]. In a neutral medium only the bubble diagram contributes, while in a charged medium both diagrams should be considered [26-30]. In the one-loop approximation, the mass operator $\Sigma(x,x')$ determines the radiative correction to the neutrino energy in the form

$$\Delta E = \frac{1}{T}\int d^4x\, d^4x'\, \overline{\psi}_\nu(x) \Sigma(x,x') \psi_\nu(x'), \qquad (1)$$

where T is the interaction time, $\psi_\nu(x)$ is the neutrino bispinor in the zero approximation without taking into account the radiative corrections.

The AMM of a lepton induced by an external field is determined by the terms in the radiative energy shift that are proportional to $(s^\mu \tilde{F}_{\mu\nu} p^\nu)$ [34]:

$$\text{Re}(\Delta E^s_H) = \frac{(\Delta\mu)(s^\mu \tilde{F}_{\mu\nu} p^\nu)}{\varepsilon_\nu}, \qquad (2)$$

where $\varepsilon_\nu = \sqrt{\mathbf{p}^2 + m_\nu^2}$ is the neutrino energy, $F_{\mu\nu} = \partial_\mu A_\nu - \partial_\nu A_\mu$, $A_\mu$ is an external electromagnetic field potential, $\tilde{F}_{\mu\nu}$ is the dual electromagnetic field tensor, $s^\mu$ is the particle polarization 4-vector defined as [34]

$$s^\mu = [\frac{\boldsymbol{\zeta} \cdot \mathbf{p}}{m_\nu}, \boldsymbol{\zeta} + \frac{\mathbf{p}(\boldsymbol{\zeta} \cdot \mathbf{p})}{m_\nu(\varepsilon_\nu + m_\nu)}], \qquad (3)$$

and $\boldsymbol{\zeta}$ is twice the mean value of the neutrino spin vector in the rest frame. Below we shall use the temperature (Matsubara) Green's functions method, which is also known as the imaginary time technique. For further presentation, it will suffice to mention that the diagram technique in the imaginary time formalism framework is analogous to the Feynman rules in the ordinary quantum field theory. The diagram technique in the imaginary time representation was developed according to the following replacements [35-36]:

$$p_0 \to i\omega_\ell + \mu, \int \frac{dp_0}{2\pi} \to iT \sum_{\ell=-\infty}^{\infty}, \qquad (4)$$

where $\mu$ is the chemical potential of a macroscopic system at a finite temperature $T = \beta^{-1}$, $\omega_n = 2\pi T(n+1/2), n = 0, \pm 1, \pm 2,...$ for fermions, and $\omega_n = 2\pi n T$ for bosons. We present here analytical expressions for the temperature Green's function of an ideal electron-positron gas in a constant magnetic field. The electron Green's function in a homogeneous and stationary magnetic field in the proper-time representation is given by the formula [37-38]

$$G(x', x'') = \Phi(x', x'') \int \frac{d^4 p}{(2\pi)^4} \exp(-ipx) G(p), \qquad (5)$$

where

$$G(p) = i \int_0^\infty ds \exp\left[-is(m^2 - p_0^2 + p_3^2 + p_\perp^2 \frac{tgz}{z})\right]\left\{\frac{m + \gamma^0 p^0 - \gamma^3 p^3}{\cos z}\exp(iz\Sigma_3) - \frac{(\boldsymbol{\gamma} \cdot \boldsymbol{p})_\perp}{\cos^2 z}\right\}, \quad (6)$$

$$\Phi(x', x'') = \exp[-ie \int_{x'}^{x''} A^{ext}_\mu(x) dx^\mu], \quad z = -eHs, \quad x = x'' - x', \quad (\boldsymbol{\gamma} \cdot \boldsymbol{p})_\perp = \gamma^1 p^1 + \gamma^2 p^2$$

and $q = -e < 0$ is the charge of the electron.

The temperature Green's function is obtained from formulas (5) and (6) by replacing

$$\int \frac{dp_0}{2\pi} \exp[-i(p_0 x_0 - \boldsymbol{p} \cdot \boldsymbol{x})] G(p_0, \boldsymbol{p}) \to iT \sum_\ell \exp[-i(\tau\omega_\ell - i\mu\tau) + i\boldsymbol{p} \cdot \boldsymbol{x}] G(p_0 = i\omega_\ell + \mu, \boldsymbol{p}), \quad (7)$$

with $\tau = \tau'' - \tau' \in [-1/T, 1/T]$.

The contribution of the W boson to the mass operator of the neutrino in the one-loop approximation is determined by the expression [39-40]

$$\Sigma^W(x, x') = -i\frac{g^2}{8}(1 - \gamma^5)\gamma^\mu G(x, x')\gamma^\nu(1 + \gamma^5) D_{\mu\nu}(x, x'), \quad (8)$$

where $G(x, x')$ and $D_{\mu\nu}(x, x')$ are the electron and W-boson propagators in a magnetic field specified by the potential

$$A^\mu = (0, 0, xH, 0). \quad (9)$$

In the Feynman gauge the W-boson propagator has the form [38, 41- 42]

$$D_{\mu\nu}(x, x') = \frac{1}{(2\pi)^2} \int_0^\infty \frac{dt}{t^2} D_{\mu\nu}(t) \exp(-iM_W^2 t) \frac{bt}{\sin(bt)} \exp\left\{-\frac{i}{4}[x_\parallel^2 - \frac{bx_\perp^2}{tg(bt)} - 4\Omega]\right\}, \quad (10)$$

where we have introduced the notation

$$\Omega = -b(x_2 - x_2')(x_1 + x_1'), \quad x_\perp^2 = (x_1 - x_1')^2 + (x_2 - x_2')^2, \quad b = eH, \quad (11)$$

and in the matrix $D_{\mu\nu}$ the following elements are nonzero:

$$D_{00} = -D_{33} = 1, \quad D_{22} = D_{11} = -\cos(2y), \quad D_{12} = -D_{21} = -\sin(2y), \quad y = bt. \quad (12)$$

Accounting for formulas (1) and (5)-(12) we obtain the following exact result for the charged weak current contribution to the massive neutrino energy shift in the magnetized medium

$$\Delta E_\nu(iE_\ell) = -\frac{g^2 T}{2\varepsilon_\nu (2\pi)^3} \sum_n d^3k \int_0^\infty ds \int_0^\infty dt \frac{1}{\cos z \cos y} \exp\{-it[M^2_W + k^2_3 + k^2_n + \frac{tgy}{y} k^2_\perp] -$$

$$-is[m^2 + (p^3 - k^3)^2 + (E_\ell - k_n)^2 + \frac{tgz}{z}(\bm{p} - \bm{k})^2]\}$$

$$\{\frac{1}{\cos z}\left[(\bm{p}\cdot\bm{q})_\perp - \left(\frac{(\bm{p}\cdot\bm{q})_\perp}{\varepsilon_\nu + m_\nu} <\bm{\sigma}\cdot\bm{p}> + m_\nu <\bm{\sigma}\cdot\bm{p}>_\perp\right)\right] + \cos(z + 2y)[(p_3 q_3) - \frac{(p_3 q_3)<\bm{\sigma}\cdot\bm{p}>}{\varepsilon_\nu + m_\nu} -$$

$$-m_\nu q^3 <\sigma_3> -i(E_\ell - k_n)(\varepsilon_\nu - <\bm{\sigma}\cdot\bm{p}>)] +$$

$$+i\sin(z + 2y)[-q^3(\varepsilon_\nu - <\bm{\sigma}\cdot\bm{p}>) - i(E_\ell - k_n)(-p^3 + \frac{p^3 <\bm{\sigma}\cdot\bm{p}>}{\varepsilon_\nu + m_\nu} + m_\nu <\sigma_3>)]\} . \tag{13}$$

where $\bm{q} = \bm{p} - \bm{k}$, $k_n = 2\pi n T$, $E_l = 2\pi(l + 1/2)T - i\mu$, and we use the shortened notation $<\bm{\sigma}>$ for the average value of the Pauli matrices over the neutrino spinor.

We note that in the complex plane of the variable $p_0$, expression (13) is only defined for a discrete set of points on the imaginary axes. The analytical continuation of this value to the upper half-plane $(\text{Im } p^0 > 0)$ defines the real component of the neutrino energy shift. This analytical continuation is possible in principle, though generally speaking, it is not a trivial problem to solve. In this study, it is essential that the temperature and time Green's functions are defined in terms of the same spectral density. The Gaussian integrals over variables $\bm{k}$ in (13) can be performed, but the resulting expression is too complex to be presented in this paper. Nevertheless we shall examine some limiting cases.

Below I consider temperatures much lower than the W-boson mass. In this case we can neglect the thermal $W-$boson contribution to the neutrino energy shift, which will be suppressed by the factor $\exp(-\frac{M_W}{T})$.

First, we consider the neutrino energy shift in a pure magnetic field at $T = 0, \mu = m,$ where $m$ is the electron mass. To obtain the desired result, it is necessary to replace the summation in (13) with integration over variable $k_0$ according to (4). Then, after performing four Gaussian integrals over variables $\bm{k}$ and $k_0,$ the results can be written as follows:

$$\Delta E_\nu = \frac{g^2}{8(2\pi)^2 \varepsilon_\nu} \int_0^\infty \frac{d\rho}{\sin\rho} \int_0^1 du \exp\{-i\rho[\lambda(1-u)+\Lambda u]-i\varphi\} \cdot \quad (14)$$

$$\left[Q_\perp\left(\frac{\sin\rho u}{\sin\rho} - u\cos(\rho(u+1))\right) - m^2_\nu u\cos(\rho(u+1)) - iu\sin(\rho(u+1))Q_z\right],$$

where

$$Q_\perp = -m_\nu <\boldsymbol{\sigma}\cdot\boldsymbol{p}> + p^2_\perp\left[1 - \frac{<\boldsymbol{\sigma}\cdot\boldsymbol{p}>}{\varepsilon_\nu + m_\nu}\right], Q_z = m_\nu\left[\varepsilon_\nu <\sigma_3> - p_3\frac{<\boldsymbol{\sigma}\cdot\boldsymbol{p}>}{\varepsilon_\nu + m_\nu}\right], \quad (15)$$

$$\varphi = \frac{p^2_\perp}{eH}\left[\frac{\sin\rho(1-u)\sin\rho u}{\sin\rho} - u\rho(1-u)\right] - \frac{m^2_\nu}{eH}u\rho(1-u).$$

Results (14)-(15) coincide with the radiative energy shift of the massive Dirac neutrino in an external magnetic field and were originally published in [43]. As a consequence of (14) - (15), for comparatively weak magnetic fields, when the condition

$$H \ll H_0 = \frac{m^2}{e} \simeq 4.41\cdot 10^{13} G, \quad (16)$$

is fulfilled, the neutrino mass shift in a constant magnetic field is described by the formula

$$\Delta m_\nu \simeq -\frac{3g^2}{64\pi^2 M^2_W}(\zeta H)em_\nu, \quad (17)$$

where the spin number $\zeta = \pm 1$ describes the particle spin orientation as being either along or contrary to the magnetic field. Using formulas (2) and (17), one can obtain the static AMM of the Dirac neutrino (DN) with the mass $m_\nu$

$$\mu^0_\nu = \frac{3eG_F m_\nu}{8\pi^2\sqrt{2}} \simeq 3\cdot 10^{-19}\frac{m_\nu}{1eV}\mu_B, \quad (18)$$

where $\mu_B = \frac{e}{2m}$ is the Bohr magneton. The neutrino magnetic moment in Eq. (18) agrees with the results of Refs. [44-45].

Result (13) was further used to describe the W-boson contribution to the energy shift and AMM of the massive neutrino in magnetized electron-positron plasma.

Our goal is to calculate the complete energy shift and AMM, including the effects of the medium and linear terms in the magnetic field. To begin, we considered the energy shift in the field-free case. Using the summation formula [46]

$$\sum_n f(\omega_n) = \frac{1}{2T}\sum_i \mathrm{Res}[f(\omega) ctg \frac{\omega}{2T}, \omega_i], \qquad (19)$$

the following exact expression can be obtained for the neutrino energy shift in the electron-positron plasma when $H = 0$:

$$\Delta E_\nu(T,\mu,H=0) = \frac{g^2}{4(2\pi)^3 \varepsilon_\nu} \sum_{\varepsilon=\pm 1} \int \frac{d^3q}{\sqrt{q^2+m^2}} \left[\exp[(\sqrt{q^2+m^2}-\varepsilon\mu)/T]+1\right]^{-1} \times \qquad (20)$$

$$\times \frac{\boldsymbol{p}\cdot\boldsymbol{q} - \dfrac{\boldsymbol{p}\cdot\boldsymbol{q}}{\varepsilon_\nu + m_\nu}<\boldsymbol{\sigma}\cdot\boldsymbol{p}> - m_\nu <\boldsymbol{\sigma}\cdot\boldsymbol{q}> - \varepsilon\sqrt{m^2+q^2}(\varepsilon_\nu - <\boldsymbol{\sigma}\cdot\boldsymbol{p}>)}{M_W^2 - m^2 - m_\nu^2 - 2\boldsymbol{p}\cdot\boldsymbol{q} - 2\varepsilon\varepsilon_\nu\sqrt{m^2+q^2}}.$$

In the high-temperature region and for relativistic neutral plasma, when the conditions

$$\mu = 0, \ \varepsilon_\nu \ll \frac{M_W^2}{T}, \ m \ll T \ll M_W, \qquad (21)$$

are fulfilled, the temperature correction to the massive polarized DN is defined by the formula

$$\Delta E_\nu = -\frac{14\pi^2}{45\sqrt{2}} \frac{\varepsilon_\nu G_F T^4}{M_W^2} \left[(1-<\boldsymbol{\sigma}\cdot\frac{\boldsymbol{p}}{\varepsilon_\nu}>) - \frac{m_\nu^2}{4\varepsilon_\nu^2}\right]. \qquad (22)$$

where $\boldsymbol{p}$ is the neutrino momentum.

For the case of cold degenerate plasma, one obtains the following result from Eq. (20):

$$\Delta E_\nu = \frac{1}{\sqrt{2}} n_e G_F [1-<\boldsymbol{\sigma}\cdot\frac{\boldsymbol{p}}{\varepsilon_\nu}>]. \qquad (23)$$

In this equation, the density $n_e$ of the free electron gas is related to the chemical potential $\mu$ by the formula

$$\mu = \sqrt{m^2 + (3\pi^2 n_e)^{2/3}}. \qquad (24)$$

Result shown in (23) is valid when the conditions

$$\varepsilon_\nu \ll \frac{M_W^2}{\mu}, \ T \ll E_F = \mu(T=0), \qquad (25)$$

are fulfilled.

In comparison with the Wolfenstein formula for the energy shift of an electron neutrino moving in a dense electron medium in the absence of an external field, the novelty of our result (23) includes the dependence from other neutrino parameters such as the neutrino spin, energy and momentum. If $m_\nu = 0,$ Eq. (23) coincides with the Wolfenstein formula for the energy shift of a massless electron neutrino in a dense electron medium.

Further we shall consider the neutrino energy shift in the case of a weak magnetic field, when the condition

$$2eH \ll \mu^2 - m^2 \qquad (26)$$

is satisfied simultaneously with the conditions shown in (16). First we consider neutrinos propagating through a magnetized, charged medium, where the chemical potential that shows the asymmetry between the particles and antiparticles is $\mu \neq 0$ and conditions (25) are also fulfilled. The pure magnetic field contribution to the energy shift is obtained immediately from Eq. (13) and is accounted for by formula (19):

$$\Delta E_\nu = -\frac{eg^2 \sqrt{\mu^2 - m^2}}{(4\pi)^2 \varepsilon_\nu M_W^2} \left\{ (\boldsymbol{p} \cdot \boldsymbol{H})(1 - <\boldsymbol{\sigma} \cdot \frac{\boldsymbol{p}}{\varepsilon_\nu}>) - \left[ \frac{(\boldsymbol{p} \cdot \boldsymbol{H})<\boldsymbol{\sigma} \cdot \boldsymbol{p}>}{\varepsilon_\nu + m_\nu} + m_\nu <\boldsymbol{\sigma} \cdot \boldsymbol{H}> - (\boldsymbol{p} \cdot \boldsymbol{H})<\boldsymbol{\sigma} \cdot \frac{\boldsymbol{p}}{\varepsilon_\nu}> \right] \right\}. \qquad (27)$$

The case of a magnetized neutral medium should also be explored when conditions (16) and (21) are satisfied. In the approximation considered, the following result can be derived from (13):

$$\Delta E_\nu = \frac{eg^2 T^2}{48 M_W^4} \left\{ -[m_\nu <\boldsymbol{\sigma} \cdot \boldsymbol{H}> + \frac{(\boldsymbol{p} \cdot \boldsymbol{H})<\boldsymbol{\sigma} \cdot \boldsymbol{p}>}{\varepsilon_\nu + m_\nu} - (\frac{\boldsymbol{p}}{\varepsilon_\nu} \cdot \boldsymbol{H})<\boldsymbol{\sigma} \cdot \boldsymbol{p}>] + 2(\boldsymbol{p} \cdot \boldsymbol{H})(1 - <\boldsymbol{\sigma} \cdot \frac{\boldsymbol{p}}{\varepsilon_\nu}>) \right\} \qquad (28)$$

Using the relation

$$m_\nu <\boldsymbol{\sigma} \cdot \boldsymbol{H}> + \frac{(\boldsymbol{p} \cdot \boldsymbol{H})<\boldsymbol{\sigma} \cdot \boldsymbol{p}>}{\varepsilon_\nu + m_\nu} - (\frac{\boldsymbol{p}}{\varepsilon_\nu} \cdot \boldsymbol{H})<\boldsymbol{\sigma} \cdot \boldsymbol{p}> = -\frac{m_\nu}{\varepsilon_\nu}(s^\mu \tilde{F}_{\mu\nu} p^\nu) \qquad (29)$$

we obtain the following expressions for the full W-contributions to the massive electron neutrino energy shift for neutral and charged magnetized plasma cases, respectively:

$$\Delta E^N{}_\nu = \frac{g^2 eT^2}{48 M^4_W}\left\{ m_\nu \frac{(s_\mu \tilde{F}^{\mu\nu} p_\nu)}{\varepsilon_\nu} + 2(\boldsymbol{p}\cdot\boldsymbol{H})(1- <\boldsymbol{\sigma}\cdot\frac{\boldsymbol{p}}{\varepsilon_\nu}>)\right\} - \frac{28\pi^2 \varepsilon_\nu G_F T^4}{45\sqrt{2} M^2_W}\left[(1- <\boldsymbol{\sigma}\cdot\frac{\boldsymbol{p}}{\varepsilon_\nu}>) - \frac{m^2_\nu}{4\varepsilon^2_\nu}\right] +$$

$$+\mu^0{}_\nu \frac{(s_\mu \tilde{F}^{\mu\nu} p_\nu)}{\varepsilon_\nu}, \quad \varepsilon_\nu \ll \frac{M^2_W}{T}, m \ll T \ll M_W, eH \ll m^2, \tag{30}$$

$$\Delta E^C{}_\nu = -\frac{g^2 e\sqrt{\mu^2-m^2}}{(4\pi)^2 \varepsilon_\nu M^2_W}\left\{ m_\nu \frac{(s_\mu \tilde{F}^{\mu\nu} p_\nu)}{\varepsilon_\nu} + (\boldsymbol{p}\cdot\boldsymbol{H})(1- <\boldsymbol{\sigma}\cdot\frac{\boldsymbol{p}}{\varepsilon_\nu}>)\right\} + \frac{1}{\sqrt{2}} n_e G_F [1- <\boldsymbol{\sigma}\cdot\frac{\boldsymbol{p}}{\varepsilon_\nu}>] +$$

$$+\mu^0{}_\nu \frac{(s_\mu \tilde{F}^{\mu\nu} p_\nu)}{\varepsilon_\nu}, \quad \varepsilon_\nu \ll \frac{M^2_W}{\mu}, T \ll E_F, 2eH \ll \mu^2 - m^2, m^2 \tag{31}$$

By accounting for Eqns. (2), (30), (31) we find that the W-contributions to the plasma correction of the neutrino AMM has the asymptotes:

$$(\Delta\mu_\nu)^N = \frac{4}{9}\pi^2 (\frac{T}{M_W})^2 \mu^0{}_\nu, \quad \varepsilon_\nu \ll \frac{M^2_W}{T}, \quad m \ll T \ll M_W, \quad eH \ll m^2, \tag{32}$$

$$(\Delta\mu_\nu)^C = -\frac{4}{3}\frac{(3\pi^2 n_e)^{1/3}}{\varepsilon_\nu} \mu^0{}_\nu, \quad \varepsilon_\nu \ll \frac{M^2_W}{\mu}, \quad T \ll E_F, \quad 2eH \ll \mu^2 - m^2. \tag{33}$$

The separate consideration shows that the contribution of the charged scalar to the neutrino energy shift and AMM is negligibly small ($\propto (m/M_W)^2$) compared to the $W$-contribution, and therefore we omit it here.

### 3. Neutral weak current contribution to the energy shift

The tadpole contribution to the mass operator of the neutrino is determined by the expression

$$\Sigma^{tadpole}(p) = -i\left(\frac{g}{2\cos\theta_w}\right)^2 R\gamma_\mu iZ^{\mu\nu}(0) \int \frac{d^4 q}{(2\pi)^4} Sp\left[\gamma_\nu (C_V + C_A \gamma^5) iG^F(q)\right], \tag{34}$$

where $Z^{\mu\nu}(0)$ is the vacuum propagator of the neutral $Z$-boson at zero momentum, $G^F(q)$ is the lepton propagator given by Eq. (6), and the quantities $C_V = -\frac{1}{2} + 2\sin^2\theta_w$ and $C_A = -\frac{1}{2}$ are the vector and axial-vector coupling constants, which come in the neutral-current interaction of the electrons, $\gamma^5 = -i\gamma^0 \gamma^1 \gamma^2 \gamma^3, R = \frac{1}{2}(1-\gamma^5)$. Using the imaginary time technique, the tadpole contribution to the self-energy in the case of a weak magnetic field becomes

$$\Delta E_\nu^{tadpole} = \Delta E_\nu^{tadpole}(H=0) + \Delta E_\nu^{tadpole}(H \neq 0), \tag{35}$$

where

$$\Delta E_\nu^{tadpole}(H=0) = \frac{G_F C_V}{\sqrt{2}} \left[1 - \frac{<\boldsymbol{\sigma}\cdot\boldsymbol{p}>}{\varepsilon_\nu}\right]\left[N_{e^-} - N_{e^+}\right], \tag{36}$$

$$\Delta E_\nu^{tadpole}(H \neq 0) = \frac{g^2 e C_A}{4M_W^2 \varepsilon_\nu}\left[\frac{m_\nu}{\varepsilon_\nu}(s\tilde{F}p) + (\boldsymbol{p}\cdot\boldsymbol{H})\left(1 - \frac{<\boldsymbol{\sigma}\cdot\boldsymbol{p}>}{\varepsilon_\nu}\right)\right]b(T,\mu). \tag{37}$$

In the expressions shown above, the net number density of the electrons (positrons) and parameter $b(T,\mu)$ are defined as

$$N_{e^\pm} = 2\int \frac{d^3q}{(2\pi)^3}\left[\exp((\sqrt{q^2+m^2}\pm\mu)/T)+1\right]^{-1}, \tag{38}$$

$$b(T,\mu) = \int \frac{d^3q}{(2\pi)^3 E}\frac{d}{dE}\left[n_F\left(\frac{E-\mu}{T}\right) - n_F\left(\frac{E+\mu}{T}\right)\right]. \tag{39}$$

It follows from (37) that the neutral weak current contribution to the AMM for each type of neutrino in a magnetized medium is described by the formula

$$(\Delta\mu_\nu)^{tadpole} = \frac{g^2 e m_\nu C_A}{4M_W^2 \varepsilon_\nu}b(T,\mu). \tag{40}$$

Thus, for $\nu_\mu$ and $\nu_\tau$ the AMM is defined by the formula (40), and in the $CP-$symmetric plasma, the neutral weak current does not contribute to the neutrino dispersion and AMM. We note that the neutral weak current does not contribute to the vacuum AMM of a neutrino either.

In the limiting case (29) for a completely degenerate electron gas, the field dependent contribution of the neutral weak current to the self-energy and AMM of a Dirac neutrino moving in weak magnetic field is determined by the asymptotes

$$\Delta E_\nu^{tadpole}(H \neq 0) = \frac{(\Delta\mu)^{tadpole}}{\varepsilon_\nu}(s\tilde{F}p) - \frac{eg^2 C_A \sqrt{\mu^2-m^2}}{(4\pi)^2 M_W^2 \varepsilon_\nu}(\boldsymbol{p}\cdot\boldsymbol{H})\left(1 - \frac{<\boldsymbol{\sigma}\cdot\boldsymbol{p}>}{\varepsilon_\nu}\right), \tag{41}$$

$$(\Delta\mu_\nu)^{tadpole} = -\frac{eg^2 m_\nu C_A}{(4\pi)^2 M_W^2}\frac{\sqrt{\mu^2-m^2}}{\varepsilon_\nu} = C_A(\Delta\mu)^W. \tag{42}$$

The plasma contribution to the AMM of an electron neutrino, determined by the summation of weak charged and neutral current contributions, is defined by the formulas (32)-(33), (40), and (42). Our results for W boson contribution in (32), which were obtained using the method of a finite temperature (Matsubara) Green's function, coincides with the corresponding results of the papers [47-48] and differs by factor of ½ from the result published in [49]. The sum of the corrections in (33) and (42) agrees with the result from [48] for charged plasma correction to the AMM of an electron neutrino in a weak magnetic field.

It should be emphasized that if the tested neutrino moves in a medium containing a neutrino background from the same family, there is a diagram similar to tadpole diagram, but with a $Z-\nu$ loop, that also contributes to the neutrino dispersion. The corresponding contribution to the energy shift may be obtained from (22) using the following replacements [24]

$$g^2 \to \frac{g^2}{2\cos^2\theta_W}, \quad M_W \to M_Z.$$

For the $CP-$symmetric medium

$$\Delta E^Z_\nu = -\frac{7\pi^2 \varepsilon_\nu G_F T^4}{45\sqrt{2} M^2_Z}\left[(1-\langle\boldsymbol{\sigma}\cdot\boldsymbol{p}\rangle\frac{1}{\varepsilon_\nu}) - \frac{m^2_\nu}{4\varepsilon^2_\nu}\right], \qquad (43)$$

where $M_Z$ is the $Z-$boson mass. As can be observed from a comparison, in the case of neutral plasma, the $Z-$contribution (43) is on the same order of magnitude as the $W-$contribution (22). By accounting for results (30) and (43) in a comparatively weak magnetic field, the full energy shift of an electron neutrino in the neutral plasma is determined by the formula

$$\Delta E_\nu = -\frac{g^2 e T^2}{48 M^4_W}\left\{m_\nu \frac{(p_\mu \tilde{F}^{\mu\nu} s_\nu)}{\varepsilon_\nu} - 2(\boldsymbol{p}\cdot\boldsymbol{H})(1-\langle\boldsymbol{\sigma}\cdot\boldsymbol{p}\rangle\frac{1}{\varepsilon_\nu})\right\} - \qquad (44)$$

$$-\frac{14\pi^2 \varepsilon_\nu G_F T^4}{45\sqrt{2} M^2_W}(1+\frac{M^2_W}{2M^2_Z})\left[(1-\langle\boldsymbol{\sigma}\cdot\boldsymbol{p}\rangle\frac{1}{\varepsilon_\nu}) - \frac{m^2_\nu}{4\varepsilon^2_\nu}\right] + \mu^0_\nu \frac{(s_\mu \tilde{F}^{\mu\nu} p_\nu)}{\varepsilon_\nu},$$

The corresponding result in the case of charged-asymmetric electron-positron plasma is determined by summing the charged weak current and neutral weak current contributions defined by formulas (31) and (36)-(37), respectively:

$$\Delta E_\nu = \frac{g^2 e(1+C_A)}{4\varepsilon_\nu M_W^2}\left\{m_\nu \frac{(s_\mu \tilde{F}^{\mu\nu} p_\nu)}{\varepsilon_\nu}+(\boldsymbol{p}\cdot\boldsymbol{H})(1-\langle\boldsymbol{\sigma}\cdot\boldsymbol{p}\rangle\frac{1}{\varepsilon_\nu})\right\}b(T,\mu)+ \tag{45}$$

$$+\frac{G_F(1+C_V)}{\sqrt{2}}\left[1-\frac{<\boldsymbol{\sigma}\cdot\boldsymbol{p}>}{\varepsilon_\nu}\right]\left[N_{e^-}-N_{e^+}\right]+\mu^0_\nu\frac{(s_\mu \tilde{F}^{\mu\nu}p_\nu)}{\varepsilon_\nu},\ 2eH \ll \mu^2-m^2, m^2.$$

In the limiting case, when $m_\nu=0, \frac{\langle\boldsymbol{\sigma}\cdot\boldsymbol{p}\rangle}{|\boldsymbol{p}|}=-1$, results (44) and (45) coincides with the energy shift of a massless left-handed neutrino in the magnetized electron-positron plasma obtained in [28-29].

### 3. Neutrino dispersion in a CP-symmetric magnetized medium: real time method and strong magnetic field.

The presence of strong magnetic fields would be expected to influence the propagation of neutrinos in the early universe producing an imprint in neutrino oscillations at those epochs. The only tool to distinguish between different neutrino flavors is the analysis of bubble impact. Therefore, with regard to magnetic field effects on neutrino oscillations, the bubble contribution is the crucial factor.

As it was already mentioned, in the $CP$ symmetric plasma the field dependent contribution on the energy shift of an electron neutrino is the $W-$ contribution. The interaction of a neutrino with a neutrally magnetized medium in this section is described based on the mass operator (1) constructed from the real time Green's function of an electron-positron plasma at finite temperature and density in a constant magnetic field [39,50-51]. We also neglect thermal corrections to the $W-$ propagator and use the vacuum propagator defined by the formulas (10)-(12). Using the explicit form of the propagators and Eq. (l), we represent the neutrino-energy shift in the form

$$\Delta E_\nu = \Delta E_\nu(H, T=0, \mu=m) + \Delta E_\nu(H, T\neq 0, \mu\neq m). \tag{46}$$

where $\Delta E_\nu(H, T=0, \mu=m)$ is the radiative energy shift of the neutrino in a magnetic field at zero temperature and zero density of the medium. This energy shift, as mentioned above, was investigated in detail in Ref. [43] and is omitted in this paper. The second term in Eq. (46) accounts for the effects of interest that arise from the finite temperature and density of the medium and is determined by the expression

$$\Delta E_\nu(H, T \neq 0, \mu \neq m) = i\frac{g^2}{32(2\pi)^2}$$

$$\sum_{n,\varepsilon=\pm 1} \int_{-\infty}^{+\infty} dk_3 \int_0^\infty dt \frac{\varepsilon(eH)}{\exp[(E_n - \varepsilon\mu)/T] + 1} \exp\{it[(q_0 - \varepsilon E_n)^2 - k_3^2 - M_W^2 - 2eHn] - i\frac{q_\perp^2}{2eH}\sin(2y)\}$$

$$\left[ -2i\varepsilon \frac{\sqrt{2eHn}}{E_n} AB I_{n,n-1} + A^2 \exp(iy)(1 + \varepsilon\frac{p_3}{E_n}) I_{n,n} + B^2 \exp(-iy)(1 - \varepsilon\frac{p_3}{E_n}) I_{n-1,n-1} \right]. \quad (47)$$

In this formula, $I_{nm}(z)$ is a Laguerre function [52] with argument

$$z = \frac{2q_\perp^2}{eH}\sin^2(y), \quad y = eHt,$$

$k_3 = q_3 - p_3$ is the z-component of the momentum of the intermediate W boson, $q = (\varepsilon_\nu, \boldsymbol{q})$ – is the four-momentum of the neutrino, the coefficients A and B, which account for the polarization state of the neutrino, are:

$$A = 2[(1 + s\frac{m_\nu}{\varepsilon^\perp_\nu})(1 - \frac{q_3}{\varepsilon_\nu})]^{1/2}, \quad B = 2[(1 - s\frac{m_\nu}{\varepsilon^\perp_\nu})(1 + \frac{q_3}{\varepsilon_\nu})]^{1/2}, \quad (48)$$

where $\varepsilon^\perp_\nu = \sqrt{\varepsilon_\nu^2 - q_z^2}$ and the spin number $s = \pm 1$ describes the neutrino spin orientation along or contrary to the magnetic field $\boldsymbol{H}$. In (47), a summation is carried out over the positive $(\varepsilon = +1)$ and negative $(\varepsilon = -1)$ frequency states of the Dirac equation solution in a constant magnetic field, and the electron energy levels in a stationary magnetic field are determined by the formula [53]

$$E_n = \sqrt{2eHn + m^2 + p_3^2},$$

where $n = 0, 1, 2, ...$ is the principle quantum number, $p_3(-\infty < p_3 < +\infty)$ is the electron momentum projection on the magnetic field direction $\boldsymbol{H}$. The number $n$ is known as the Landau level number [54]. We chose the transverse-polarization operator $\mu_3$ as the operator determining the polarization state of the Dirac neutrino and plasma particles [52]:

$$\mu_3 = \Sigma_3 + \rho_2 \frac{[\Sigma, \mathbf{P}]_3}{m}, \quad P^\mu = i\frac{\partial}{\partial x_\mu} - qA^\mu.$$

Here

$$\Sigma_k = \begin{pmatrix} \sigma_k & 0 \\ 0 & \sigma_k \end{pmatrix}, \quad \rho_2 = \begin{pmatrix} 0 & -iI \\ iI & 0 \end{pmatrix}$$

and $\sigma_k$ are the Pauli matrices.

Next we perform the summation in Eq. (47) over the principle quantum number $n$ using the formula [52]

$$\sum_{n=0}^{\infty} \exp(-2i\beta s n) u_n(\eta) u_n(\eta') = \frac{\exp(i\beta s)}{(2\pi i \sin(2\beta s))^{1/2}} \exp\left[-i\frac{\eta\eta'}{\sin(2\beta s)} + \frac{i}{2}\text{ctg}(2\beta s)(\eta^2 + \eta'^2)\right], \quad (49)$$

where $u_n(\eta)$ is a Hermit function.

As a result, the $W$ – contribution to the electron neutrino energy shift in the electron-positron plasma at temperature $T$ and chemical potential $\mu$ in an external field $\mathbf{H}$ pointing in the z direction is written in the form

$$\Delta E_\nu = \frac{g^2}{32(2\pi)^3} \sum_{\varepsilon=\pm 1} \int_{-\infty}^{+\infty} dp_3 \int_{-\infty}^{+\infty} d\tau \int_0^\infty dt \int_0^\infty d\lambda \frac{eH\varepsilon}{\exp[(E-\varepsilon\mu)/T]+1} \frac{1}{\sin(\beta\tau)}$$

$$\exp\left[i\tau\lambda + it(q_\perp^2 + m^2 + \lambda + m_\nu^2 - M_W^2 - 2\varepsilon q_0 E + 2p_3 q_3) - i\frac{q_\perp^2 \sin(\beta\tau)\sin(\beta t)}{eH\sin(\tau+t)\beta}\right] \quad (50)$$

$$\left\{A^2(1+\varepsilon\frac{p_3}{E})\exp[i\beta(\tau+2t)] + B^2(1-\varepsilon\frac{p_3}{E})\exp[-i\beta(\tau+2t)] + 2\varepsilon\frac{ABq_\perp \sin(\beta t)}{E\sin\beta(\tau+t)}\right\},$$

where $\beta = eH, E = \sqrt{m^2 + \lambda + p_3^2}$. In the following section, results (47) and (50) are investigated in various limiting cases.

A further simplification of Eq. (50) occurs if the neutrino moves in the direction perpendicular to the magnetic field, when $q_3 = 0$. In this case the plasma contribution to the radiative energy shift of the neutrino can be represented as the sum of two parts:

$$\Delta E_\nu(H, T \neq 0, \mu \neq m) = \Delta E_1(s, H, T, \mu) + \Delta E_2(H, T, \mu), \qquad (51)$$

where the part of the energy shift, $\Delta E_1(s, H, T, \mu)$, that depends explicitly on the orientation of the neutrino spin and the part, $\Delta E_2(H, T, \mu)$, of the energy shift that does not depend on the orientation of the neutrino spin are respectively

$$\begin{pmatrix} \Delta E_1 \\ \Delta E_2 \end{pmatrix} = \frac{g^2}{4(2\pi)^3} \sum_{\varepsilon=\pm 1} \int_{-\infty}^{+\infty} dp_3 \int_{-\infty}^{+\infty} d\tau \int_0^\infty dt \int_0^\infty d\lambda \frac{\varepsilon}{\exp[(E-\varepsilon\mu)/T]+1} \frac{eH}{\sin(\beta\tau)} \qquad (52)$$

$$\begin{pmatrix} is\frac{m_\nu}{\varepsilon_\nu}\sin(\beta(\tau+t)) \\ \cos(\beta(\tau+t)) + 2\varepsilon\frac{q_\perp^2 \sin(\beta t)}{E\varepsilon_\nu \sin(\beta\tau)} \end{pmatrix} \exp\left[ i\tau\lambda + it(q_\perp^2 + m^2 + m_\nu^2 - M_W^2 - 2\varepsilon q_0 E) - i\frac{q_\perp^2 \sin(\beta t)\sin(\beta(\tau-t))}{eH \sin(\tau\beta)} \right]$$

It should be emphasized that the full dependence of the radiative energy shift from the neutrino spin number $s$ in the case when the neutrino is moving perpendicular to the magnetic field is determined only by the part of the energy shift that is explicitly dependent on the neutrino spin orientation [49], i.e.

$$\Delta E_1 = -sH(\Delta\mu_\nu) \qquad (53)$$

where $\Delta\mu_\nu$ is the neutrino AMM in the magnetized electron-positron plasma.

Let us first to find the asymptotes for the values $\Delta E_1$ and $\Delta E_2$ in Eqs. (52) for comparatively weak magnetic fields, when the following conditions

$$H \ll H_0 = \frac{m^2}{e} \simeq 4{,}41 \cdot 10^{13} \text{Гc}. \qquad (54)$$

are fulfilled.

In a CP-symmetric plasma, when the conditions

$$\mu = 0, \; m \ll T \ll M_W, \; q_\perp \ll \frac{M_W^2}{T}, \qquad (55)$$

are also fulfilled, the contribution on the electron neutrino energy shift in the field – free case and the AMM asymptotes are described by the formulas

$$\Delta E_2 = -\frac{7\pi^2}{90}\left(\frac{T}{M_W}\right)^4 g^2 \varepsilon_\nu, \qquad (56)$$

$$\Delta\mu_\nu(T,\mu=0) \simeq \frac{em_\nu g^2}{48 M^2_W}\left(\frac{T}{M_W}\right)^2 = \frac{4}{9}\pi^2\left(\frac{T}{M_W}\right)^2 \mu^0_\nu, \qquad (57)$$

In the limiting case of a degenerate electron gas, when the following conditions are fulfilled

$$T \ll \mu = E_F,\ 2eH \ll \mu^2 - m^2,\ m^2,\ \varepsilon_\nu \ll \frac{M^2_W}{\mu}, \qquad (58)$$

it follows from Eq. (52) that

$$\Delta E_2(H=0, T=0, n_e \neq 0) = \sqrt{2} n_e G_F, \qquad (59)$$

$$\Delta\mu_\nu = -\frac{\Delta E_1}{sH} = -\frac{4}{3}\frac{(3\pi^2 n_e)^{1/3}}{\varepsilon_\nu}\mu^0_\nu. \qquad (60)$$

Notice that result (59) coincides with the Wolfenstein formula for the energy shift of an electron neutrino moving in a dense electron medium in the absence of an external field, and it has no dependence on the spin orientation.

It should be noticed that the early universe was almost charge symmetric, with a particle-antiparticle asymmetry of only $10^{-9} - 10^{-10}$ [17, 55].

Let us first investigate the neutrino dispersion in the neutral medium with a strong magnetic field under the conditions

$$\mu = 0,\ \sqrt{eH} \gg T \gg m. \qquad (61)$$

Such a strong magnetic field can be expected to exist in the early universe before the period of neutrino decoupling era, which is believed to occur around the energy scale of 1 MeV when the age of the universe was approximately 1 second [30, 32].

Because in the strong-field limit (61) the gap between the electron Landau levels is larger than the temperature, only the lowest Landau level contributions in the formula (47) of the plasma contribution to the neutrino energy shift should be used to be consistent.

From Eq. (47) in the LLL approximation [30], the asymptote for the electron neutrino energy shift in a $CP$- symmetric electron-positron plasma in a strong magnetic field takes the form

$$\Delta E_\nu = -\frac{g^2(eH)\varepsilon_\nu T^2}{48 M_W^4}(1+s\frac{m_\nu}{\varepsilon_\nu}). \quad (62)$$

It follows from (62) that under the conditions shown in (61) the plasma contribution to the neutrino AMM is determined by the formula (57) obtained in the weak field approximation.

We also note that in a superstrong magnetic field, the CP-symmetric plasma contribution (62) to the neutrino energy greatly exceeds the similar correction (22) in the field-free case:

$$\left|\frac{\Delta E_\nu(62)}{\Delta E_\nu(22)}\right| \simeq \frac{eH}{(2\pi T)^2}. \quad (63)$$

To estimate the energy shift of the neutrino we set $(M_W/T)=50, (eH/T^2)=250, \varepsilon_\nu=10^2 MeV$. Then from Eq. (22) and (62) we have $|\Delta E_\nu(22)| \simeq 4.8 eV$ and $|\Delta E_\nu(62)| \simeq 30 eV$. For $(M_W/T)=100$, $(eH/T^2)=10^3$, $\varepsilon_\nu=10^2 MeV$ one can obtain that $|\Delta E_\nu(22)| \simeq 0.3 eV$ and $|\Delta E_\nu(62)| \simeq 7.5 eV$ respectively. As follows from the estimates given in the case (61) the energy shift of the neutrino can be substantially greater than its value in the magnetic field free case.

In the limiting case of a degenerate electron gas and relatively strong magnetic fields, when

$$T \ll E_F = \mu(T=0), \quad M_W^2 \gg eH > (\mu^2-m^2)/2, \quad (64)$$

the first term of the expansion of (48) in the parameter $(T/E_F) \ll 1$ corresponds to replacing the Fermi distribution function by a $\theta$-function and the energy shift and AMM of the neutrino are determined by Eq. (47). In this case it is only necessary to take the term n=0 in the sum. It follows from (64) that for $H \sim (10^{-6}-10^2)H_0$ at $n_e < 10^{19}-10^{31} cm^{-3}$ only the ground level can be occupied with electrons. For example, in modern models of neutron stars, the star core (with its thickness ~ 0.1 of its radius) appears as a crystal lattice of the ions immersed in a highly degenerate gas of relativistic electrons with an electron density $n_e \leq 10^{38} \, cm^{-3}$, temperature $T \sim 10^6 - 10^7 \, K$ and magnetic field strength $H \sim 10^{12} - 10^{14} \, G$ [56-57]. Fields of the order of $10^{16} \, G$ and larger could exist in the magnetars [30, 58]. The typical energy of neutrinos emitted via URCA processes and at the conditions of a supernova core is about $1 \, MeV$ and $10^2 \, MeV$, respectively.

If we require also that the condition $\varepsilon_\nu \ll \dfrac{M^2_W}{\mu}$ be fulfilled, and taking into account the relationship between the chemical potential and the density of the electron gas [56]

$$p_F = 2\pi^2 \frac{n_e}{eH}, \qquad (65)$$

we find the following asymptotes in the first approximation

$$\mathrm{Re}(\Delta E_\nu) \simeq \frac{g^2 eH}{16\pi^2}(1+\zeta \frac{m_\nu}{\varepsilon_\nu})\frac{p_F}{M^2_W},$$

$$\Delta \mu_\nu = -\frac{8\pi^2}{3}\frac{n_e}{(eH)\varepsilon_\nu}\mu^0_\nu. \qquad (66)$$

To estimate numerical values of the obtained correction (66) to the static AMM of a neutrino in the limiting case (64) we set

$$n_e \simeq (1/2)\pi^{-2} m^3 \left(\frac{H}{H_0}\right)^{3/2},$$

which corresponds to a chemical potential value given by $p_F = m\left(\dfrac{H}{H_0}\right)^{1/2}$. Then from Eq. (66) we have $\dfrac{|\Delta\mu_\nu|}{\mu^0_\nu} = \dfrac{4m}{3\varepsilon_\nu}\left(\dfrac{H}{H_0}\right)^{1/2}$. For $H = H_0$, $\varepsilon_\nu = 1 MeV$ and $\varepsilon_\nu = 10 MeV$ one can find $\dfrac{|\Delta\mu_\nu|}{\mu^0_\nu} \approx 0.7$ and $\dfrac{|\Delta\mu_\nu|}{\mu^0_\nu} \approx 0.07$ respectively. Assuming, that $H = 10^2 H_0$, $\varepsilon_\nu = 1 MeV$ and $\varepsilon_\nu = 10 MeV$ we obtain that $\dfrac{|\Delta\mu_\nu|}{\mu^0_\nu} \approx 7$ and $\dfrac{|\Delta\mu_\nu|}{\mu^0_\nu} \approx 0,7$ respectively.

Now let us suppose that the neutrino moves parallel (or antiparallel) to the magnetic field. From Eq. (50) it follows that in this case the neutrino energy shift can be determined by the following exact formula:

$$\Delta E_\nu = \frac{g^2}{8(2\pi)^3} \sum_{\varepsilon=\pm 1} \int_{-\infty}^{+\infty} dp_3 \int_{-\infty}^{+\infty} d\tau \int_0^\infty dt \int_0^\infty d\lambda \frac{\varepsilon \exp(i\lambda\tau)}{exp[(E-\varepsilon\mu)/T]+1} \frac{eH}{\sin(eH\tau)} \exp[-it(M^2+2\varepsilon q_0 E - 2p_3 q_3)]$$

$$\left[(1+s)(1-\frac{q_3}{\varepsilon_\nu})(1+\varepsilon\frac{p_3}{E})\exp(i\beta(\tau+t))+(1-s)(1+\frac{q_3}{\varepsilon_\nu})(1-\varepsilon\frac{p_3}{E})\exp(-i\beta(\tau+t))\right]. \quad (67)$$

The structure of equation (67) is in agreement with the results obtained when a neutrino moves along the magnetic field; its chirality is conserved and it coincides with the spin polarization along the field, given by the operator $\mu_3$.

In a comparatively weak magnetic field and for neutral relativistic plasma, when conditions (54) and (55) are fulfilled, the energy shift (64) in the field free case takes the form

$$\Delta E_\nu(H=0) = -\frac{7\pi^2 g^2}{240} \varepsilon_\nu \left(\frac{T}{M_W}\right)^4 \left(1-s\frac{q_3}{\varepsilon_\nu}\right)\left(1-s\frac{1}{3}\frac{q_3}{\varepsilon_\nu}\right). \quad (68)$$

The pure magnetic field contribution to the energy shift is determined by the asymptotes

$$\Delta E_\nu(H\neq 0) = \frac{g^2 eH}{48 M_W^2} T^2 \left[2q_3(1-s\frac{q_3}{\varepsilon_\nu})-s\frac{m_\nu^2}{\varepsilon_\nu}\right]. \quad (69)$$

Finally we shall discuss the connection between the neutrino dispersion and AMM if the neutrino moves parallel (or antiparallel) to the magnetic field. To give the correct interpretation for result (69), we notice that it can be written in the form

$$\Delta E_\nu(H\neq 0) = \frac{g^2 eT^2}{48 M_W^4}\left[2(\boldsymbol{q}\cdot\boldsymbol{H})(1-\frac{sq_3}{\varepsilon_\nu})+m_\nu \frac{s^\mu \tilde{F}_{\mu\nu} q^\nu}{\varepsilon_\nu}\right], \quad (70)$$

where $s^\mu$ is the neutrino polarization 4-vector defined by the formula (3). The first terms in (70) in the limiting case, when $m_\nu = 0$, coincides with the energy shift of a massless left-handed neutrino in the magnetized electron-positron plasma obtained in [28-29]. The AMM of a neutrino is determined by the term in the energy shift that is proportional to $s^\mu \tilde{F}_{\mu\nu} p^\nu$ and coincides with the results in (57) obtained for a neutrino moving perpendicular to a magnetic field.

## 4. Conclusions

We argue that the results of this work represent an advance on previous studies [24-33] examining the neutrino dispersion in the magnetized medium. Our calculations may also have practical applications in the modern studies of cosmology and supernova physics, which investigate the possible existence of massive sterile neutrinos. At present, the nonrelativistic sterile neutrinos with masses on the order of 2 to 50 keV are treated as the most popular candidates for particles that form dark matter [7-8, 12-15]. It is clear, however, that the presented calculations could not be simply applied to massive sterile neutrinos. This is a separate problem that depends on the mixing scheme, type and magnetic moment (transition or diagonal) of the neutrino and will be considered elsewhere.

**Acknowledgments**

The author is grateful to V.Ch. Zhukovsкii, A.E.Shabad, A.V. Borisov and A.I. Ternov for useful discussions.